\documentclass[twocolumn,superscriptaddress,showpacs,aps,amsmath,amssymb]{revtex4}

\usepackage{amssymb}
\usepackage{mathrsfs}
\usepackage{epsfig}
\usepackage{amsbsy}
\usepackage{amsmath}
\usepackage{amsfonts}
\usepackage{graphicx}
\usepackage{latexsym}

\begin{document}
\title{Decoherence suppression of open quantum systems through a strong coupling to non-Markovian reservoirs}
\author{Chan U Lei}
\affiliation{Department of Physics, California Institute of
Technology, Pasadena, CA 91125}
\author{Wei-Min Zhang}
\email{wzhang@mail.ncku.edu.tw} \affiliation{Department of Physics
and Center for Quantum Information Science, National Cheng Kung
University, Tainan 70101, Taiwan }

\begin{abstract}
In this paper, we provide a mechanism of decoherence suppression for
open quantum systems in general, and that for "Schrodinger cat-like"
state in particular, through the strong couplings to non-Markovian
reservoirs. Different from the usual strategies of suppressing
decoherence by decoupling the system from the environment in the
literatures, here the decoherence suppression employs the strong
back-reaction from non-Markovian reservoirs. The mechanism relies on
the existence of the singularities (bound states) of the
nonequilibrium retarded Green function which completely determines
the dissipation and decoherence dynamics of open systems. As an
application, we examine the decoherence dynamics of a photonic
crystal nanocavity that is coupled to a waveguide. The strong
non-Markovian suppression of decoherence for the optical cat state
is attained.
\end{abstract}

\date{Oct. 30, 2011}

\pacs{03.67.Pp; 03.65.Yz; 42.50.Dv; 05.70.Ln}

\maketitle

\section{Introduction}

How to protect quantum states away from decoherence is one of the
most challenge topics in quantum information processing and modern
quantum technology. During the past two decades, many schemes have
been theoretically proposed  and experimentally realized to suppress
the decoherence in quantum information processing \cite{Feedback,
DFS, DynDecoupling,Yao07077602,UDD07100504,Du091265}. On the other
hand, due to the significant development of the nanotechnology
during the past decade, various quantum devices with high
tunabilities, such as nanomechanical oscillator or superconducting
qubit strongly coupled to cavity \cite{CirQED, CavityMec}, trapped
atom coupled to an engineered reservoir \cite{AtomEngRes}, arrays of
coupled nanocavities in photonic crystal \cite{CavityCROW} etc., can
be engineered. In these quantum devices, the strong coupling between
the system and the structured reservoir and the resulting
non-Markovian back-action play an important role in the
manipulations of quantum coherence.

In this work, we shall provide a general mechanism of decoherence
suppression for quantum systems coupled strongly to non-Markovian
reservoirs. Contrary to the ordinary means of suppressing
decoherence via dynamically decoupling of the system from the
environment \cite{DynDecoupling,Yao07077602,UDD07100504,Du091265},
we employ the strong non-Markovian back-reaction from the
environment to suppress the decoherence of quantum states. We show
in general that when the non-Markovian back-reaction is strong
enough, the decoherence of quantum states can be largely suppressed.
In particular, we examine the time evolution of the Wigner function
for a mesoscopic superposition of two coherent states, and
demonstrate that the decoherence of such a mesoscopic superposition
state can be suppressed due to the strong non-Markovian
back-reaction from the environment.


\section{Exact Master Equation}

The dynamics of open quantum systems are described by the reduced
density matrix which can be obtained by tracing over all of the
reservoir degrees of freedom form the total system $\rho(t) =
\text{tr}[\rho_{\text{tot}}(t)]$, where $\rho_{\text{tot}}(t)$ is
the total density matrix of the system plus its reservoir. The exact
master equation of the reduced density matrix $\rho(t)$ for an open
system, such as a cavity in quantum optics, a defect (nanocavity) in
photonic crystals or a quantum dot in nanostructures, etc., coupled
to a general non-Markovian reservoir has been derived recently
\cite{Xio10012105,Wu1018407,Tu08235311,Lei}
\begin{align}
\label{masterequation} \frac{d\rho(t)}{dt} = & \frac{1}{i}[H'(t),
\rho(t)] + \gamma(t)[
2a\rho(t)a^{\dag} - \rho(t) a^{\dag}a - a^{\dag}a\rho(t)] \notag \\
&+ \tilde{\gamma}(t)[a\rho(t)a^{\dag} + a^{\dag}\rho(t)a -
a^{\dag}a\rho(t) - \rho(t)aa^{\dag}] \ ,
\end{align}
where $H'(t) = \omega'(t)a^{\dag}a$ is the renormalized Hamiltonian
of the system with the renormalized frequency $\omega'(t) = -
\text{Im}[\dot{u}(t)u^{-1}(t)]$. The time-dependent coefficients
$\gamma(t) = - \text{Re}[\dot{u}(t)u^{-1}(t)]$ and
$\tilde{\gamma}(t) = \dot{v}(t) -
2v(t)\text{Re}[\dot{u}(t)u^{-1}(t)]$ incorporates all of the
dissipations and fluctuations induced from the coupling to the
reservoir. The function $u(t)$ is the nonequilibrium retarded Green
function of the system satisfying the following equation:
\begin{equation}
\label{ueq} \dot{u}(t) + i \omega_c u(t) + \int_{t_0}^{t} g(t-\tau)
u(\tau) = 0
\end{equation}
subjected to the initial condition $u(t_0) = 1$, and the
nonequilibrium thermal fluctuation is characterized by the function
$v(t)$ which is given by
\begin{equation}
\label{vsol} v(t) = \int_{t_0}^t d\tau \int_{t_0}^t d\tau'
u^*(\tau_1)\tilde{g}(\tau_1 - \tau_2) u(\tau_2).
\end{equation}
By introducing the spectral density of the reservoir $J(\omega)=2\pi
\sum_k |V_k|^2 \delta(\omega - \omega_k)$ where $V_k$ is the
coupling between the system and the reservoir, the time correlation
functions $g(\tau-\tau')$ and $\tilde{g}(\tau-\tau')$ in
Eqs.~(\ref{ueq}-\ref{vsol}) are given by
\begin{align}
\label{timecorrel-1} & g(\tau-\tau') =
\int_0^{\infty} \frac{d\omega}{2\pi}J(\omega)e^{-i\omega(\tau-\tau')} \ , \\
\label{timecorrel-2} & \tilde{g}(\tau-\tau') =
\int_0^{\infty}\frac{d\omega}{2\pi}J(\omega)\bar{n}(\omega, T)
e^{-i\omega(\tau-\tau')} \ ,
\end{align}
which characterize all the non-Markovian back-reactions of the
reservoir, and $\bar{n}(\omega, T) = \frac{1}{e^{\hbar\omega/k_B T}
- 1}$ is the average particle number distribution in the reservoir
at the initial time $t_0$.

The decoherence dynamics of quantum states can be studied by
examining the evolution of the corresponding Wigner function. With
the help of the exact master equation (\ref{masterequation}), the
exact Wigner function of an arbitrary quantum state at arbitrary
time $t$ in the complex space $\{z \}$ is found:
\begin{equation}
\label{WignerDef} W(z, t) = \int d\mu(\alpha_0) d\mu(\alpha'_0)
\langle \alpha_0 | \rho(t_0) | \alpha'_0 \rangle
\mathfrak{J}(z,t|\alpha_0, \alpha'^{*}_0, t_0) ,
\end{equation}
where $|\alpha\rangle = e^{\alpha a^{\dag}} |0\rangle$ is the
coherent state, $d\mu(\alpha) = \frac{d\alpha^* d\alpha}{2\pi i}
e^{-|\alpha|^2}$ is the integral measure of the Bergmann complex
space, $\rho(t_0)$ is the reduced density matrix of the initial
state, and the propagating function $\mathfrak{J}(z,t|\alpha_0,
\alpha'^{*}_0, t_0)$ is given by
\begin{align}
\label{WignerPropagatingND} \mathfrak{J}(z, t \ |& \alpha_0,
\alpha'^{*}_0, t_0) = W_0 (z, t) \exp \big\{ z^* \Omega(t) u(t)
\alpha_0 \notag \\ &  + \alpha'^*_0 u^*(t) \Omega(t) z + \alpha'^*_0
\big( 1 - |u(t)|^2 \Omega(t) \big)\alpha_0
 \big\}\ ,
\end{align}
where $\Omega(t) = \frac{2}{1+v(t)}$ and $W_0(z, t) =
\frac{\Omega(t)}{\pi}\exp{[\Omega(t) |z|^2]}$.

To concentrate on quantum decoherence, we examine the time evolution
of a mesoscopic superposition of two coherent states moving in
opposite directions, called as the "Schrodinger cat-like" state or
the optical cat state in the literature \cite{SCS}: $|\phi\rangle =
N(|\alpha\rangle + |-\alpha\rangle)$, where $N =
1/\sqrt{4\cosh{|\alpha|^2}}$ is the normalization factor. As a
result of Eq.~(\ref{WignerDef}), the time-evolution of the Wigner
function for this cat state is given by
\begin{subequations}
\label{WignerCat}
\begin{equation}
 W(z,t) = W_{\alpha}(z,t) + W_{-\alpha}(z,t) + W_I(z,t)
\end{equation}
with
\begin{align}
\label{WignerPeak} & W_{\pm\alpha}(z,t) = N^2\frac{\Omega(t)}{\pi}
e^{|\alpha|^2} e^{
\Omega(t)|z \mp u(t)\alpha|^2 } \ , \\
\label{WignerInterference} W_I(z, t) & = 2N^2\frac{\Omega(t)}{\pi}
e^{-|\alpha|^2} \Re \big\{ e^{ -[z - u(t)\alpha]^* \Omega(t) [z +
u(t)\alpha]} \big\} \ .
\end{align}
\end{subequations}
In Eq.~(\ref{WignerCat}), the first two terms are the Wigner
functions for the initial coherent states $|\alpha\rangle$ and
$|-\alpha\rangle$ respectively, the third term is the interference
between them. The quantum coherence of the cat state can then be
characterized by the fringe visibility function
\begin{align}
\label{FV} F(\alpha,t) & \equiv \frac{1}{2}
\frac{W_I(z,t)|_{\text{peak}}}{\sqrt{W_{\alpha}(z,t)|_{\text{peak}}W_{-\alpha}(z,t)|_{\text{peak}}}}
\notag \\ & = \exp{\Big\{-2|\alpha|^2 (1 -
\frac{|u(t)|^2}{1+2v(t)})\Big\}}
\end{align}
which ranges from unity to $\exp{(-2|\alpha|^2  )}$ for full
coherence to complete decoherence. As shown by Eqs.~(\ref{vsol}),
(\ref{WignerCat}) and (\ref{FV}), the nonequilibrium retarded Green
function $u(t)$ completely determine the dynamics of the quantum
decoherence of the system. Equation (\ref{ueq}) alone can also give
the exact solution of atomic systems involving only single
excitation (single photon process) at zero temperature
\cite{Gar972290,Bre02}.

\section{General mechanism of decoherence suppression}

The solution of the retarded Green function can be obtained by the
inverse Laplace transformation \cite{Cohen,Kofman,Longhi} $u(t) =
\frac{1}{2\pi i} \int_B ds \tilde{u}(s) e^{st}$, where $\tilde{u}(s)
= \frac{i}{is - \omega_c - \Sigma(s)}$ and the Bromwich path B is a
line $\text{Re}(s) = \text{const} > 0$ in the half plane of the
analyticity of the transformation.  The self-energy $\Sigma(s) =
\int \frac{d\omega}{2 \pi} \frac{J(\omega)}{is - \omega}$ is the
Laplace transformation of the correlation function
(\ref{timecorrel-1}).  Consider a spectral density ranged from
$\omega_e$ to infinity, e.g. $\omega_e=0$ for Ohmic, super-Ohmic and
sub-Ohmic reservoirs, etc. The self-energy is then not defined on
the segment of the imaginary axis $s = -i \omega$ with $\omega >
\omega_e$, while $s = -i \omega_e$ is a branch point. Near the
imaginary axis, the self-energy function can be separated into real
and imaginary parts by the relation $\lim_{\eta \rightarrow 0}
\frac{1}{\omega \pm i\eta} = {\cal P}\frac{1}{\omega} \mp i \pi
\delta(\omega)$ so that $\Sigma(s = -i\omega \pm 0^+) =
\Delta(\omega) \mp i \frac{J(\omega)}{2}$ with $\Delta(\omega) =
{\cal P} \int^\infty_{\omega_e} \frac{d\omega'}{2\pi}
\frac{J(\omega')}{\omega - \omega'}$, and $\rm{\cal P}$ denotes the
principal value. The analytic properties of the transformed retarded
Green function $\tilde{u}(s)$ determine completely the decoherence
dynamics of the system.

In the very weak coupling regime, the self-energy function is
dominated near the pole $s = -i\omega_c$. The functions
$\Delta(\omega)$ and $J(\omega)$ can be approximated by $\Delta_c =
\Delta(\omega_c)$ and $J_c = J(\omega_c)$. The resulting retarded
Green function becomes $u(t) = e^{-i\omega'_c - \frac{J_c}{2}t}$
with the shifted frequency $\omega'_c = \omega_c + \Delta_c$. The
retarded Green function experiences an exponential decay with the
decay constant $J_c/2$, which reproduces the Born-Markov result
\cite{Xio10012105, Lei}. Thus,  $u(t)$ will eventually be damped to
zero and the fringe visibility will be decayed to
$\exp{(-2|\alpha|^2)}$, namely the quantum coherence is totally
lost.

However, as the coupling increases, the variation of the self-energy
away from the pole $-i\omega_c$ becomes significant, and the
decoherence dynamics of the system is then totally different from
the Born-Makov limit. Especially, there exists an isolated pole $s
=-i\Omega$ on the imaginary axis outside the branch cut, i.e.
$\Omega - \omega_c = \Delta(\Omega)$ with $\Omega<\omega_e$, which
leads to a dissipationless dynamics of the system.
\begin{figure}
\centering
\includegraphics[width = 8 cm]{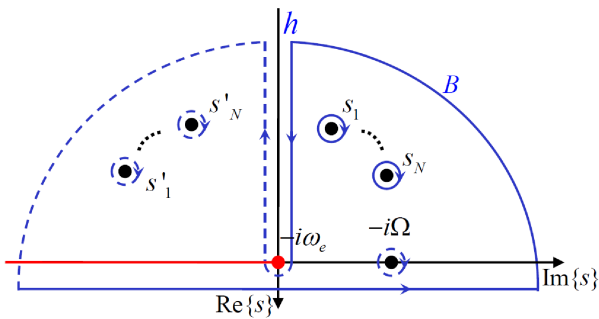}
\caption{Integration contour of the inverse laplace transform of
$\tilde{u}(s)$. The Red line on the imaginary axis is the branch
cut. $-i\Omega$ is the pure imaginary pole. The integration along
the solid (dashed) curve is made on the first (second) Riemannian
sheet. $s_i$ and $s'_i$ are the poles of $\tilde{u}(s)$ on the first
and second Riemannian sheet in the $\text{Re}(s) < 0$ half-plane.}
\label{contour}
\end{figure}
The exact solution of the retarded Green function can be obtained by
the inverse Laplace transform along the Bromwich path B as shown in
Fig.~\ref{contour}. Since the closure crosses the branch cut
$\text{Im}(s) < -\omega_e$ on the imaginary axis, the contour is
necessary to pass into the second Riemannian sheet in the section of
the half plane with $\text{Im}(s) < -\omega_e$, where it remains in
the first Riemannian sheet in the sections $\text{Im}(s) > -\omega_e
$ in the half plane $\text{Re}(s) < 0$. To properly close the
contour, it is necessary to turn around the branch point $-i
\omega_e$, following the Hankel paths $h$ to enter and leave the
second Riemannian sheet, as shown in Fig.~\ref{contour}. The exact
propagating function can be obtained by means of the residue method
\begin{align}
\label{exactu} u(t) = & \mathcal{Z} e^{-i\Omega t} +  \sum_i
\mathcal{Z}_i e^{-\gamma_i t-i\omega_i t} + \sum_i \mathcal{Z}'_i
e^{-\gamma'_i t-i\omega'_i t} \notag \\ & + \frac{1}{2\pi i}
\int_{-\infty - i\omega_e}^{0 - i\omega_e} ds \big[
\tilde{u}^{\text{II}}(s) - \tilde{u}^{\text{I}}(s) \big] e^{st} \ ,
\end{align}
where $\mathcal{Z}$ is the residues of the bound state with the
imaginary pole $s = - i \Omega$, $\mathcal{Z}_i$($\mathcal{Z}'_i$)
is the residues of the $i$th unstable states with the pole $ s = s_i
= -\gamma_i - i \omega_i$ ($s'_i = -\gamma'_i - i \omega'_i$) on the
first (second) Riemannian sheet which is the solution of $is -
\omega_c - \Sigma^{\text{I},\text{II}}(s) = 0$ with
$\Sigma^{\text{II}}(s) = \Sigma^{\text{I}}(s) + i J(i s)$. The last
term is the contribution from the contour along the Hankel path $h$
(Fig.~\ref{contour}) which is responsible to the nonexponential
decay dynamics \cite{Cohen}. As a result [shown by
Eq.~(\ref{exactu})], the retarded Green function shows the
dissipationless dynamics due to the existence of the bound state.
This means that the decoherence of the system can be suppressed
through the strong non-Markovian coupling to a reservoir. The
coherence preservation in the cat state is also obvious by
substituting Eq.~(\ref{exactu}) into Eq.~(\ref{FV}). It is
straightforward to extend the above analysis to structured
reservoirs with finite spectrum, as shown explicitly in the
following discussion.



\section{An example for applications}

\begin{figure}
\centering
\includegraphics[width = 9 cm]{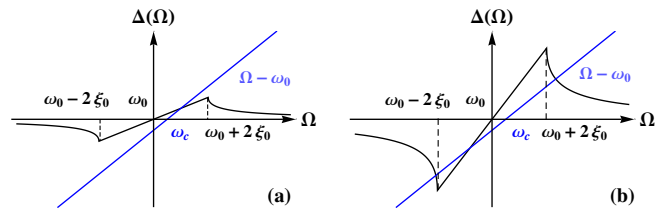}
\caption{Graphical solutions of the imaginary poles of the retarded
Green function. (a) Below the critical coupling, no solution outside
the branch cut. (b) Over the critical coupling, two solutions
outside the branch cut. } \label{boundcond}
\end{figure}

As an application, we apply the above general mechanism to the
decoherence dynamics of a nanocavity (with frequency $\omega_c$)
coupled to a structured waveguide [its characteristic dispersion
$\omega_k = \omega_0 - 2\xi_0 \cos{(k)}$]. The coupling strength
between the nanocavity and the waveguide in photonic crystals is
$V_k = \sqrt{\frac{2}{\pi}} \xi \sin{(nk)}$ \cite{Longhi,Wu1018407}.
The spectral density $J(\omega)$ is then given by
\begin{align}
\label{spectraldensity} J(\omega) = \left\{
\begin{array}{ll}
\eta^{2}\sqrt{4\xi_0^{2} - (\omega - \omega_0)^{2}}
 \ , & |\omega - \omega_0| \leq 2 \xi_0 \  \\
0 \ , & |\omega - \omega_0| > 2 \xi_0
\end{array} \right.
\end{align}
where $\eta = \xi/\xi_0$ characterizes the strength of the coupling
between the nanocavity and the structured reservoir. From the above
spectral density, the self-energy in $\tilde{u}(\omega)$ can be
exactly calculated
\begin{equation}
\label{selfenergy} \Sigma(s) = \frac{i}{2}\eta^2 \big[ (s +
i\omega_0) - \sqrt{(2\xi_0)^2 + (s + i\omega_0)^2} \ \big] \ .
\end{equation}
As the coupling strength exceeds the critical value $\eta_c =
\sqrt{2-\frac{|\omega_c - \omega_0|}{\xi_0}}$, bound modes (the
poles determined graphically in Fig.~\ref{boundcond}) occur. As a
result, when the coupling strength is below the critical coupling,
no imaginary pole exists outside the branch cut, see
Fig.~\ref{boundcond}(a). The solution of the retarded Green function
shows a dissipative dynamics. However, when the coupling strength is
larger than the critical coupling, one or two imaginary poles
appear, see Fig.~\ref{boundcond}(b), and the solution of $u(t)$
behaves dissipationless after a short time.

\begin{figure} \centering
\includegraphics[width = 9.5 cm]{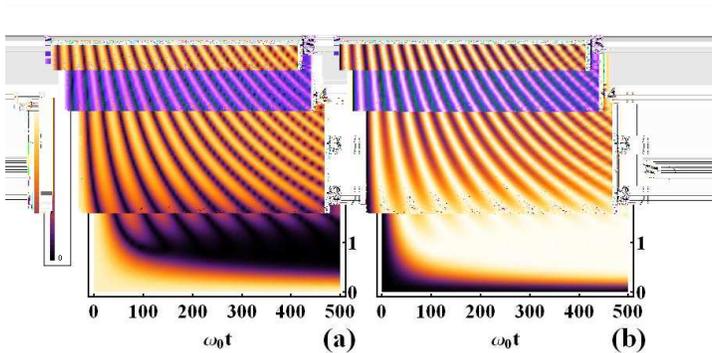}
\caption{(a) The exact numerical result of the retarded Green
function $u(t)$ and (b) the thermal fluctuation
$v(t)/\bar{n}(\omega_0, T)$ in different coupling strength. The
frequency of the nanocavity $\omega_c$ is set to be the same as the
band center of the waveguide $\omega_0$.} \label{uv}
\end{figure}

\begin{figure}
\centering
\includegraphics[width = 8 cm]{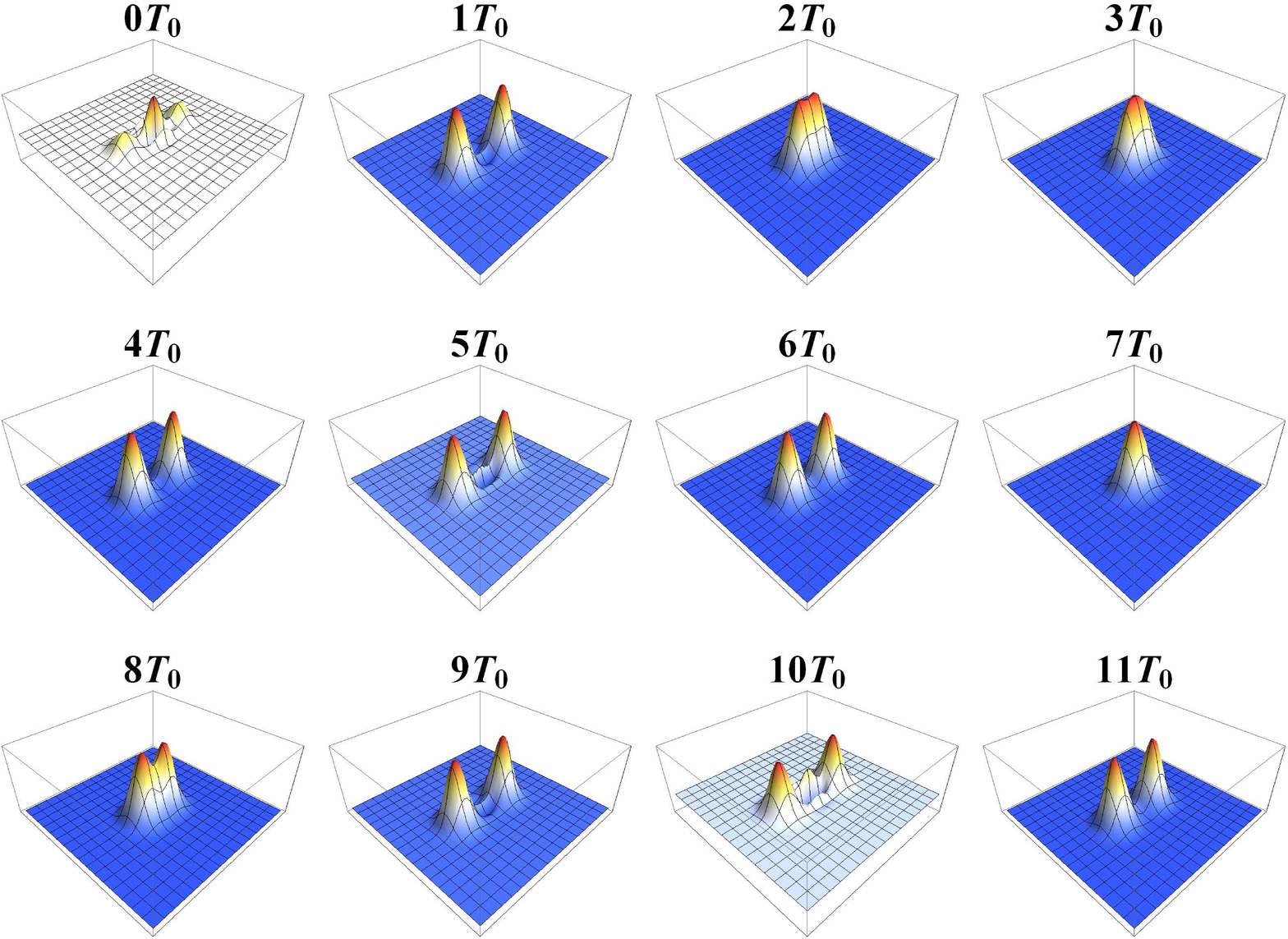}
\caption{Time evolution of the Wigner function of the mesoscopic
superposition state. The coupling strength $\eta = 4$ and the cavity
frequency equals to the band center of the waveguide $\omega_c =
\omega_0$, $T_0 = 2\pi/\omega_0$. A movie for the above
non-Markovian time evolution is given in \cite{sm}.} \label{fig4}
\end{figure}

To see explicitly the mechanism of decoherence suppression through
the strong non-Markovian effect, we may look at the steady state
solution of the nanocavity in the strong coupling regime. Consider
the case the frequency of the nanocavity equals to the band center
of the reservoir, i.e. $\omega_c = \omega_0$, the steady state
solution of $u(t)$ becomes
\begin{equation}
\label{usteady} u_{\text{st}}(t) = A(\eta)
e^{-i\omega_0t}\cos{[\omega(\eta) t ]} \ .
\end{equation}
This shows that the retarded Green function is enveloped by the
cosine function with the amplitude $A(\eta) = \frac{\eta^2 -
2}{\eta^2 - 1}$ and the frequency $\omega(\eta) =
\frac{\eta^2}{\sqrt{\eta^2-1}}\xi_0$ which corresponds to the energy
exchange between the cavity and the reservoir. Fig.~\ref{uv} shows
the exact numerical result of the retarded Green function $u(t)$
[see Fig.~\ref{uv}(a)] and the normalized thermal fluctuation
$v(t)/\bar{n}(\omega_0, T)$ [i.e. Fig.~\ref{uv}(b)] in different
coupling strength. Note that when the coupling $\eta > \eta_c =
\sqrt{2}$, both the retarded Green function $u(t)$ and the thermal
fluctuation $v(t)$ keep oscillating rather than damping.
The oscillation indicates that the cavity keeps exchanging photons
with the waveguide due to the strong non-Markovian back-reaction
from the reservoir.
The steady solution of the fringe visibility function of
Eq.~(\ref{FV}) at zero temperature simply becomes
\begin{equation}
F(\alpha,t) = \exp{\{-2|\alpha|^2(1 - A(\eta)^2 \cos^2{[\omega(\eta)
t]}\}} \ .
\end{equation}
Instead of full decoherence, the cat state keeps oscillating in the
strong coupling regime. The stronger the coupling strength is, the
larger the degree of coherence can be maintained. Fig.~\ref{fig4}
shows the periodic motion of the Wigner function for the cat state
with the coupling $\eta = 4$ and the temperature $T = 0.5$mK. As
shown in Fig.~\ref{fig4}, the interference of the cat state keeps
oscillation in time.

In fact, in the weak coupling regime, the fringe visibility will
eventually decay to $e^{-2|\alpha|^2}$ because of the decoherence
induced by the reservoir. Then all the coherence information of the
cat state will be lost. At the same time, the larger the initial
temperature of the reservoir is, the faster the decoherence
processes. According to (\ref{WignerPeak}), as the the retarded
Green function decays to zero, the two peaks of the Wigner function
gradually spiral to the origin (see the movie for this Markovian
time evolution given in \cite{sm}) and the thermal fluctuation
$v(t)$ saturates to the equilibrium value $\bar{n}(\omega_0, T)$ due
to energy relaxation. The cat state finally decays to a thermal
state with the Wigner function
\begin{equation}
W(z,t \rightarrow \infty)= \frac{2}{\pi[1+\bar{n}(\omega_0, T)]}
\exp{[-\frac{2|z|^2}{1+\bar{n}(\omega_0, T)}]} \ .
\end{equation}

In contrast, as shown in the previous analysis, when the coupling
strength exceeds the critical coupling $\eta_c$, the decoherence
dynamics of the cavity field is totally suppressed. The fringe
visibility, after a short time decay, oscillates above the value of
$e^{-2|\alpha|^2}$ for all the time. In other words, the coherence
of the cat state goes to dead and birth repeatedly. In addition,
according to Eq.~(\ref{WignerPeak}) and the stationary solution of
Eq.~(\ref{usteady}) in the strong coupling regime, the two peaks of
the Wigner function would keeps spiralling in and out of the origin
with the frequency $\frac{\eta^2}{\sqrt{\eta^2 - 1}}\xi_0$ due to
the energy exchange between the system and the reservoir, see the
movie for this non-Markovian time evolution in \cite{sm}. Thus, the
cavity field would never be thermalized by the reservoir and the
decoherence of the system is significantly suppressed.

\section{Conclusion}

In conclusion, we have shown through the exact master equation that
the nonequilibrium retarded Green function can completely determine
decoherence dynamics. From analytic properties of the retarded Green
function, we provided a general mechanism of decoherence suppression
through the strong non-Markovian back-reaction from environments. In
particular, when the coupling between the system and the reservoir
exceeds a critical coupling, the bounded modes (the imaginary poles
of the retarded Green function) leads to a dissipationless dynamics
such that decoherence can be largely suppressed, as a strong
non-Markovian memory effect. This generic behavior is explicitly
demonstrated through the decoherence dynamics of the cat state.
Since the nonequilibrium retarded Green function is well-defined for
arbitrary open quantum system, the mechanism presented in this work
should also be applicable to other more complicated open systems.

\begin{acknowledgements}
This work is supported by the National Science Council (NSC) of ROC
under Contract No. NSC-99-2112-M-006-008-MY3. We also acknowledge
the support from the National Center for Theoretical Science of NSC.

\end{acknowledgements}

\end{document}